\documentclass[twocolumn,showpacs,superscriptaddress,preprintnumbers,amsmath,amssymb]{revtex4}
\usepackage{graphicx}% Include figure files
\usepackage{dcolumn}% Align table columns on decimal point
\usepackage{bm}% bold math

\begin{document}

\title{Thermodynamic properties and structural stability of thorium dioxide}
\author{Yong Lu}
\affiliation{LCP, Institute of Applied Physics and Computational
Mathematics, Beijing 100088, China}

\author{Yu Yang}
\affiliation{LCP, Institute of Applied Physics and Computational
Mathematics, Beijing 100088, China}

\author{Ping Zhang}
\thanks{Author to whom correspondence should be
addressed. E-mail: zhang\_ping@iapcm.ac.cn} \affiliation{LCP,
Institute of Applied Physics and Computational Mathematics, Beijing
100088, China} \affiliation{Center for Applied Physics and
Technology, Peking University, Beijing 100871, China}

\date{\today}% It is always \today, today,
             %  but any date may be explicitly specified

\begin{abstract}

Using density functional theory (DFT) calculations, we have
systematically investigated the thermodynamic properties and
structural stabilities of thorium dioxide (ThO$_2$). Based on the
calculated phonon dispersion curves, we calculate the thermal
expansion coefficient, bulk modulus, and heat capacities at
different temperatures for ThO$_2$ under the quasi-harmonic
approximation. All the results are in good agreement with
corresponding experiments proving the validity of our methods. Our
theoretical studies can help people more clearly understand the
thermodynamic behaviors of ThO$_2$ at different temperatures. In
addition, we have also studied possible defect formations and
diffusion behaviors of helium in ThO$_2$, to discuss its structural
stability. It is found that in intrinsic ThO$_2$ without any Fermi
energy shifts, the interstitial Th$_i^{4+}$ defect other than oxygen
or thorium vacancies, interstitial oxygen, and any kinds of Frenkel
pairs, is most probable to form with an energy release of 1.74 eV.
However, after upshifting the Fermi energy, the formation of the
other defects also becomes possible. For helium diffusion, we find
that only through the thorium vacancy can it happen with the small
energy barrier of 0.52 eV. Otherwise, helium atoms can hardly
incorporate or diffuse in ThO$_2$. Our results indicate that people
should prevent upshifts of the Fermi energy of ThO$_2$ to avoid the
formation of thorium vacancies and so as to prevent helium caused
damages.

\end{abstract}


%\keywords{Suggested keywords}%Use showkeys class option if keyword
                              %display desired
\maketitle

\section{INTRODUCTION}

As the world's demands for energy keep growing, corresponding
researches on developing new energy sources or enhancing the
energy-consuming efficiency are attracting more and more attentions.
Until now, the fissile nuclear reactor is still a very important
energy source, in which uranium dioxide (UO$_2$) has been the main
fuel component for many years
\cite{Belle69,Killeen80,Kudin02,Dorado09,Dorado10}. However, during
the burning cycle of UO$_2$, considerable amounts of redioactive
elements emerge in the reaction waste \cite{Petit10}. And such
radioactive waste results in very troublesome long-term storage
requirements. Based on these facts, many efforts have been done to
look for possible substitutions of UO$_2$. Thorium based materials,
which are naturally abundant, are identified as good candidates for
replacing UO$_2$ in fissile nuclear reactors because they are able
to produce fewer transuranic (TRU) compared to uranium- and
plutonium-based fuels. And although $^{233}$Th is not fissile, it
can absorb a slow neutron and then form fissile $^{233}$U just
undergoing multiple beta-decays \cite{Herring}. Moreover, because of
the good solubility between thorium dioxide (ThO$_2$) and other
transuranic dioxides, using the mix oxides (MOX) of thorium and
plutonium in nuclear reactors can also help reducing the large
plutonium stockpile while maintaining acceptable safety and control
characteristics of the reactor system \cite{Tommasi}. In comparison
with previous uranium-based fuels, the thorium-based fuels also have
many additional physical advantages, such as higher melting points,
higher corrosion resistivity, lower thermal expansion coefficients,
and higher thermal conductivity \cite{Maitura}. In recent years, the
thorium-based fuels have already been tested in different reactors
\cite{Chang,Herring,Carrera,Juhn}.

As in the front position of actinide element, thorium and its
compounds have been studied ever since 1950. At the earliest stage,
a series of experimental measurements have been carried out on the
thermal properties of ThO$_2$ such as the thermal expansion
coefficient, heat capacity, and thermal conductivity
\cite{Momin,Springer,Weilbacher,McElroy,ARF,Murabayashi,Pillai,Bradshaw,Moore,Mufti}.
In 1997, Bakker {\it et al.} \cite{Bakker} presented a conclusion
and comparison for the measured values. On the other side, Chadwick
and Graham \cite{Chadwick}, Allen and Tucker \cite{Allen}, and Veal
\emph{et al.} \cite{Veal} investigated the valence-band structures
of thorium and its oxides by means of X-ray photoemission
spectroscopy. And the pressure-induced phase transition of ThO$_2$
has been studied by Jayaraman \cite{Jayaraman}, Dancausse
\cite{Dancausse}, and Idiri \cite{Idiri} \emph{et al.} experimently.
Recently in 2006, researchers from the international atomic energy
agency (IAEA) built a thermal-physical database of materials for
light water reactors and heavy water reactors \cite{Kim}. And the
earlier experimental measurements are re-addressed.

Despite the vast experimental measurements, it is to our surprise
that no one has ever theoretically investigated the thermodynamic
properties and defect behaviors for ThO$_2$, which are critically
important for its usage in thermonuclear reactors. Only recently,
several theoretical studies have been carried out on the mechanical
and electronic properties \cite{Kanchana,Shein,Sevik}, phase
transition behaviors \cite{Boettger,Wang}, and elastic and optical
properties \cite{Boudjemline} for ThO$_{2}$. Our previous study has
already proven that the thorium 5$f$ states is no longer localized
after electronic hybridizations, and density functional theory
calculations are enough to produce correct descriptions for the
ground-state properties of ThO$_2$ \cite{Wang}. So in our present
paper, we decide to systematically investigate the thermodynamic
properties and structural stabilities of ThO$_2$, by using density
functional theory calculations. The thermodynamic stability will be
discussed based on the calculated thermal parameters, while the
structural stability will be discussed by calculating the formation
energy of different kinds of defects, and diffusion energy barriers
of helium in ThO$_2$. The rest of the paper is organized as follows.
The computation methods are introduced in Section II. The
discussions about the thermodynamic properties and structural
stabilities of ThO$_2$ are presented in Section III. Finally, we
give our conclusions in Section IV.

\section{Calculation Methods}

The density functional theory (DFT) calculations are carried out
using the Vienna \textit{ab initio} simulations package (VASP)
\cite{G.Kresse1,G.Kresse2} with the projector-augmented-wave (PAW)
potential methods \cite{PAW}. The cutoff energy for the plane-wave
basis set is set to 500 eV. The exchange and correlation effects are
described by generalized gradient approximation (GGA) in the
Perdew-Burke-Ernzerhof (PBE) form \cite{PBE}. A 2$\times$2$\times$2
supercell is employed to study defect formation and helium
diffusions inside ThO$_2$. For calculations of the unit cell (12
total atoms) and 2$\times$2$\times$2 supercell (96 total atoms), the
integration over the Brillouin Zone is done on
13$\times$13$\times$13 and 5$\times$5$\times$5 $k$-ponit meshes
generated using the Monkhorst-Pack \cite{Monkhorst} method, which
are both proven to be sufficient for energy convergence of less than
1.0$\times$10$^{-4}$ eV per atom. During the supercell calculations,
the shape and size of the supercell are fixed while all the ions are
free to relax until the forces on them are less than 0.01 eV/\AA.

For a semiconductor, the Helmholtz free energy $F$ at volume $V$ and
temperature $T$ can be expressed as
\begin{equation}
F(V,T)=E(V)+F_{vib}(V,T)\label{eq1},
\end{equation}
where $E(V)$ stands for the ground-state electronic energy,
$F_{vib}(V,T)$ is the phonon free energy at a given unit cell volume
$V$. Within quasi-hamonic approximation (QHA), $F_{vib}(V,T)$ can be
evaluated by
\begin{equation}
F_{vib}(V,T)=k_{B}T\sum_{j,\mathbf{q}}\ln\left[ 2\sinh\left(
\frac{\hbar\omega_{j}(\mathbf{q},V)}{2k_{B}T}\right)  \right], \label{eq2}%
\end{equation}
where $\omega_{j}(\mathbf{q},V)$ is the phonon frequency of the
$j$th phonon mode with wave vector $\mathbf{q}$ at fixed $V$, and
$k_{B}$ is the Boltzmann constant. The total specific heat of the
crystal is the sum of all phonon modes over the Brillouin zone (BZ),
\begin{equation}
C_{v}(T)=\sum_{j,\mathbf{q}}c_{v,j}(\mathbf{q},T). \label{eq3}
\end{equation}
$c_{v,j}(\mathbf{q},T)$ is the mode contribution to the specific
heat defined as,
\begin{equation}
c_{v,j}(\mathbf{q},T)=k_{B}\sum_{j,\mathbf{q}}\left(\frac{\hbar\omega_{j}(\mathbf{q},V)}{2k_{B}T}\right)^{2}\frac{1}{\sinh^{2}[\hbar\omega_{j}(\mathbf{q},V)/2k_{B}T]}.\label{eq4}
\end{equation}
The mode Gr\"{u}neisen parameter $\gamma_{j}(\mathbf{q})$ describing
the phonon frequency shift with respect to the volume can be
calculated by
\begin{equation}
\gamma_{j}(\mathbf{q})=-\frac{d[\ln\omega_{j}(\mathbf{q},V)]}{d[\ln
V]}. \label{eq5}
\end{equation}
The acoustic Gr\"{u}neisen parameter $\gamma(T)$ defined as the
weighted average of the mode Gr\"{u}neisen parameters for all
acoustic phonon branches is calculated to be
\begin{equation}
\gamma(T)=\frac{\alpha_{v}(T)B(T)V_{m}(T)}{C_{v}(T)}, \label{eq6}
\end{equation}
where $\alpha_{v}(T)$ is the thermal expansion coefficient and
equals to $\frac{1}{V}\left( \frac{\partial V}{\partial
T}\right)_{P}$, $V_{m}(T)$ is the volume per mole material, and
$B(T)$ and $C_{v}(T)$ are the bulk modulus and specific heat
respectively.

The formation energy of a point defect X with charges $q$ can be
calculated by introducing the chemical potential concept as,
\begin{equation}
E_{for}({\rm X}^{q})=E_{tot}\pm n_{x}\mu -E_{\rm ThO_{2}}+q(E_{v}+E_{f}+\Delta V),%
\label{eq7}
\end{equation}
where $E_{tot}$ is the total energies of the supercell with defect
X, $n_{x}$ represents the number of X defects, $\mu$ is the chemical
potential of X with a positive sign for vacancy and a negative sign
for interstitial defect, $E_{\rm ThO_2}$ is the energy of the
ThO$_2$ supercell without defects, $E_{v}$ and E$_{f}$ are the
valence-band maximum (VBM) and the Fermi level of ThO$_2$
respectively. With these denotations, $E_{for}({\rm V}_{\rm O}^0)$,
$E_{for}({\rm O}_i^0)$, $E_{for}({\rm V}_{\rm Th}^0)$, and
$E_{for}({\rm Th}_i^0)$ represent for the formation energies of a
neutral oxygen vacancy, a neutral interstitial oxygen, a neutral
thorium vacancy, and a neutral interstitial thorium respectively.
The shift of the VBM in a defect supercell $\Delta V$ takes the
change of the valence-band maximum caused by the defect into
account. Its value can be obtained by a macroscopic average
technique \cite{Baldereschi,Peressi} through calculating the
difference between the average electrostatic potential in a bulklike
environment of the defect supercell and the average electrostatic
potential in the defect-free supercell. The formation energy of a
Frenkel-pair can be calculated by summing up the formation energies
of a vacancy and a corresponding interstitial add-in, i.e.,
\begin{equation}
E_{for}({\rm FP_{\rm X}})=E_{for}({\rm V_{\rm X}})+E_{for}({\rm X_{i}}).\label{eq8}%
\end{equation}
For the Schottky defect, the formation energy can be calculated by
\begin{equation}
E_{for}({\rm S})=E_{for}({\rm V_{\rm Th}})+2E_{for}({\rm V_{\rm O}})-\frac{3(N-1)}{N}E_{\rm ThO_{2}},\label{eq9}%
\end{equation}
where $N$ is the number of atoms in the considered supercell. In
this expression, the defect consists of a thorium vacancy and two
oxygen vacancies, which are again supposed to be non-interacting.

\section{Results and discussions}

\subsection{Structure and Elastic constants of ThO$_2$}

Our previous studies have shown that DFT calculations with GGA
exchange-correlation functionals are good enough for obtaining the
ground-state properties of ThO$_2$ \cite{Zhang10}, and adding
additional $U$ or $J$ modifications for localization effects might
lead to incorrect results. We believe that it is because the 5$f$
electronic states of Th become delocalized after electronic
hybridizations in ThO$_2$. Here based on our previous studies, we
further investigate the thermodynamic properties of ThO$_2$.

From the mechanical point of view, the theoretical equilibrium
volume $V_{0}$, bulk modulus $B_{0}$, and the pressure derivative of
bulk modulus $B'$ can be obtained by fitting the third-order
Brich-Murnaghan equation of state \cite{Brich}. In this way, our
calculated lattice parameter a$_{0}$ for ThO$_{2}$ is 5.619 \AA,
which is in accordance with the experimental data of 5.60 \AA~
\cite{Idiri,Olsen}. The bulk modulus $B_{0}$ and its pressure
derivative $B'$ are calculated to be 190 GPa and 4.3, also in
agreement with corresponding experimental values of 195-198 GPa and
4.6-5.4 \cite{Idiri,Olsen}, respectively. In order to evaluate the
Poisson$'$s ratio $\nu$, we calculated the three independent elastic
constants $C_{11}$, $C_{12}$ and $C_{44}$ of ThO$_{2}$. The
calculation methods are the same as in our previous studies
\cite{Wang,Lu,Zhang10}. The obtained elastic constants for ThO$_2$
are $C_{11}$=351.2 GPa, $C_{12}$=106.9 GPa, and $C_{44}$=74.1 GPa,
which are in accordance with experimentally measured values of
$C_{11}$=367 GPa, $C_{12}$=106 GPa, and $C_{44}$=79 GPa
\cite{Macedo}. Furthermore, the Poisson$'$s ratio $\nu$ is
calculated to be 0.293, in excellent accordance with the
experimental data of 0.285 \cite{Macedo}. All the good agreement
between our calculated values and corresponding experimental
measurements indicates that our calculation methods are effective
and reliable. In comparison with other actinide dioxides,we can see
that ThO$_2$ has a slightly smaller bulk modulus than UO$_2$ and
PuO$_2$ \cite{Zhang10}.

\subsection{Phonon dispersions, Thermal expansion, and Heat capacity of ThO$_2$}

\begin{figure}[ptb]
\begin{center}
\includegraphics[width=1.0\linewidth]{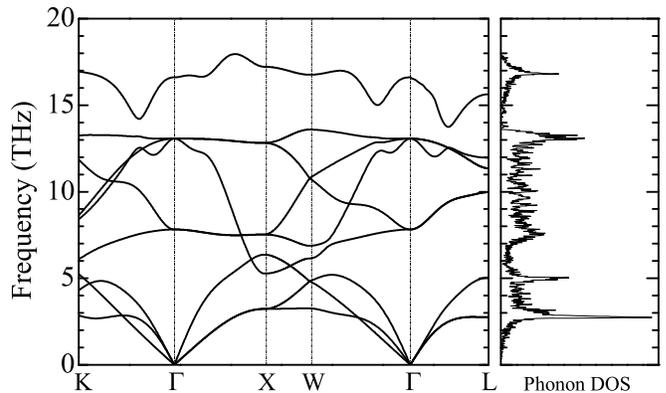}
\end{center}
\caption{(Color online) Phonon dispersion curves and phonon density
of states (DOS) for ThO$_{2}$.} \label{1}
\end{figure}

\begin{figure}[ptb]
\begin{center}
\includegraphics[width=1.0\linewidth]{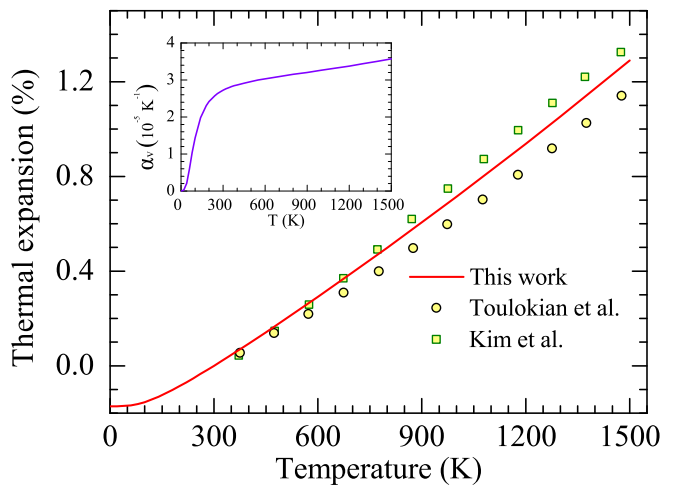}
\end{center}
\caption{(Color online) Temperature dependence of the linear thermal
expansion for ThO$_{2}$. The inset is the volume thermal expansion
coefficient as a function of temperature.}\label{2}
\end{figure}

The thermodynamic properties of a material are connected to its
phonon dispersion curves. In order to get accurate splittings
between longitudinal (LO) and transverse optical (TO) phonon
branches, the Born effective charge is firstly calculated. Because
of the high symmetry of ThO$_{2}$, the Born transverse effective
charge tensor $\mathbf{Z}^{*}$ is the same along the [100], [010],
and [001] directions, and the effective charge can be averaged by
$Z^{*}=\frac{1}{3}Tr\mathbf{Z}^{\ast}$. For ThO$_2$, we obtain that
$Z_{\mathrm{Th}}^{\ast}$=+5.41, and $Z_{\mathrm{O}}^{\ast}$=$-$2.71.
The static dielectric constant
$\epsilon=\frac{1}{3}Tr\mathbf{\epsilon}$ is 4.83. The dielectric
constant might be smaller than experimental measurements because of
the underestimation of the electronic energy band gap due to the
drawback of the exchange-correlation approximation (GGA). The
calculated phonon dispersion curves along the high-symmetry k-point
lines using the above Born effective charges are shown in Fig. 1 .
As clearly shown, there is an obvious splitting between longitudinal
optical (LO) and TO branches due to the Born effective charge.
Moreover, we find that there is no evident gap between acoustic and
optical branches of phonon for ThO$_2$, with an observable overlap
between the longitudinal acoustic (LA) and transverse optical (TO)
branches around the X point. Detailed vibrational modes analysis
tells us that the vibrations of Th and O atoms dominate the low-
(0-6 THz) and high-frequency (6-18 THz) modes, respectively.

From the obtained phonon dispersion curves and Eqs. (1) and (2), we
then calculate the energy curves for ThO$_2$, and find the
lowest-energy lattice constants at different temperatures. The
lattice expansion curve is thus obtained and shown in Fig. 2. The
experimental data by Touloukian \emph{et al.} \cite{Touloukian} and
Kim \emph{et al.} \cite{Kim} are also shown in Fig. 2 for
comparisons with our calculational results. One can see that in the
temperature range from 300 to 600 K, our result is in excellent
agreement with the experimental data. In a higher temperature range
from 600 to 1500 K, our obtained theoretical values are in the
middle of the two different experimental measurements. And at 1500
K, the relative difference between our result and the two
experimental values are 0.15\% and 0.05\%, respectively. The small
relative differences indicate that the QHA method can give
reasonable lattice parameters for ThO$_2$ up to 1500 K. The thermal
expansion coefficient $\alpha_{v}(T)$ is calculated and shown in the
inset of Fig. 2. The experimental values for the thermal expansion
coefficient in the temperature range from 298 to 1500 K can be
calculated from the corresponding thermal expansion curves, which
are 3.318$\times$10$^{-5}K^{-1}$ for Touloukian's data and 3.630
$\times$10$^{-5}K^{-1}$ for Kim's data, respectively. Our
theoretical result of 3.509 $\times$10$^{-5}K^{-1}$ is consistent
with the experimental values.

The bulk modulus $B$ is also analyzed as a function of temperature
according to the formula $B=V_{0}(\frac{\partial^{2}F}{\partial
V^{2}})_{V_{0}}$, and the result is displayed in Fig. 3. We can see
that the value of bulk modulus decreases with increasing
temperature, and at 1500 K, the ratio of bulk modulus ($B/B_{0}$) is
0.898, as depicted in the inset of Fig. 3.

\begin{figure}[ptb]
\begin{center}
\includegraphics[width=0.8\linewidth]{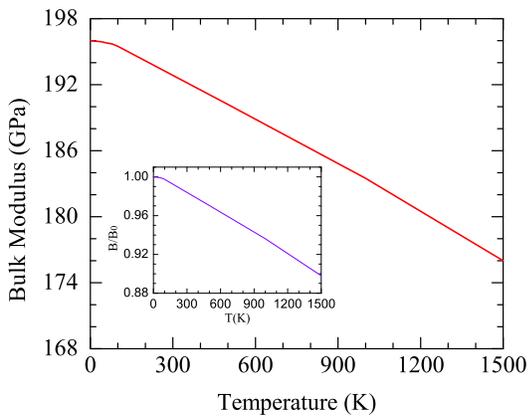}
\end{center}
\caption{(Color online) Temperature dependence of bulk modulus $B$
for ThO$_{2}$. The inset is the ratio of $B/B_{0}$.} \label{3}
\end{figure}

\begin{figure}[ptb]
\begin{center}
\includegraphics[width=0.9\linewidth]{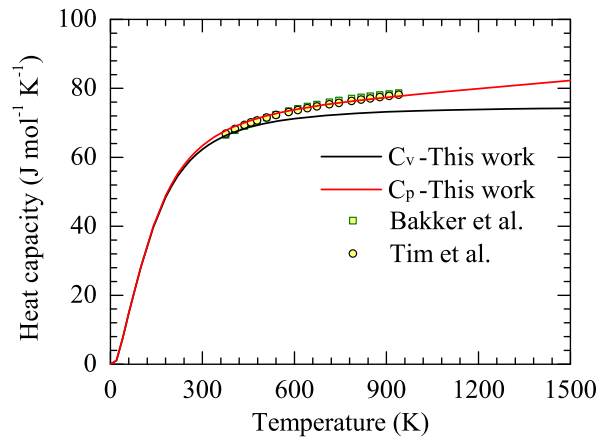}
\end{center}
\caption{(Color online) Heat capacities at constant volume ($C_{v}$)
and constant pressure ($C_{p}$) of ThO$_{2}$.} \label{4}
\end{figure}

Under QHA, the considered vibration modes are harmonic but
volume-dependent. The calculated heat capacity at constant volume
using Eqs. (\ref{eq3}) and (\ref{eq4}) is shown in Fig. 4, together
with the heat capacity at constant pressure $C_{p}$, which is
calculated according to the relationship,
\begin{equation}
C_{p}-C_{v}=\alpha_{v}^{2}(T)B(T)V(T)T.\label{eq10}
\end{equation}
The available experimental data from Bakker \emph{et al.}
\cite{Bakker} and Kim \emph{et al.} \cite{Kim} are also shown in
Fig. 4 for comparisons. It is evident that our theoretical result is
in excellent agreement with the measured values for the whole
experimentally considered temperature range. As temperature
increases, the value of $C_{p}$ increases continuously, while the
value of $C_{v}$ approaches to a constant of 3$R$ ($R$ is the
Rydberg constant). At 1500 K, the value of $C_{p}$ becomes 81
J$\cdot$mol$^{-1}$K$^{-1}$. In general, QHA method is valid when the
temperature is much lower than the material's melting point (around
3600 K for ThO$_{2}$ \cite{Bakker}) when the anharmonic effect is
small.

\subsection{Gr\"{u}neisen parameters and thermal conductivity of ThO$_2$}

The lattice thermal conductivity $\kappa_{L}$ for a material can be
calculated differently, depending on the specific mechanisms for
phonon scattering. At relative high temperatures, the dominant
mechanism for phonon scattering is the Umklapp process, in which the
acoustic phonon branches interact with each other and transport
heat. With this mechanism, the lattice thermal conductivity of a
crystal-like material can be expressed as
\cite{Slack,Slack1979,Anderson},
\begin{equation}
\kappa_{L}=A\frac{\bar{M}\Theta^{3}(T)\delta(T)n^{2/3}}{\gamma^{2}(T)\times
T}, \label{eq11}
\end{equation}
where $A$ is a physical constant with the value of
3.1$\times$10$^{-6}$, $\bar{M}$ is the average mass per atom in the
crystal, $\Theta(T)$ is the Debye temperature of ThO$_2$, $n$ is the
number of atoms in the primitive unit cell, $\gamma(T)$ is the
acoustic Gr\"{u}neisen parameter, and $\delta(T)$ is the cube root
of the average volume per atom, i.e., the averaged radius per atom.
The $\kappa_{L}$ and $\delta$ in Eq. (\ref{eq11}) are in units of W
m$^{-1}$K$^{-1}$, and \AA~ respectively. With reasonable expressions
of the Debye temperature and acoustic Gr\"{u}neisen parameter to
describe the harmonic phonon branches and the anharmonic
interactions between different phonon branches, Eq. (\ref{eq11}) can
provide accurate predictions for a material's thermal conductivity
\cite{Slack1979}.

Firstly the Debye temperature $\Theta$ can be determined from the
elastic constants within the Debye theory, in which the vibrations
of the solid are considered as elastic waves, and the Debye
temperature of the solid is related to an averaged sound velocity
\cite{Blanco}. Within isotropic approximation, the Debye temperature
$\Theta$ can be expressed as \cite{Blanco},
\begin{equation}
\Theta(T)=\frac{\hbar}{k_{B}}[6\pi^{2}V^{1/2}(T)n]^{1/3}f(\nu)\sqrt{\frac{B(T)}{M}},
\label{eq13}
\end{equation}
where M is the molecular mass per formula unit, $B(T)$ is the bulk
modulus, $\nu$ is the material's Poisson$'$s ratio, and $f(\nu)$ is
given by \cite{Francisco,Francisco2}
\begin{equation}
f(\nu)=3^{1/3} \left[2 \left(\frac{2}{3}\frac{1+\nu}{1-2\nu}
\right)^{3/2}+ \left(\frac {1}{3}\frac{1+\nu}{1-\nu} \right)^{3/2}
\right]^{-1/3}. \label{eq14}
\end{equation}
Our calculations find that the Debye temperature monotonically
decreases, while the acoustic Gr\"{u}neisen parameter monotonically
increases with increasing temperature.

\begin{figure}[ptb]
\begin{center}
\includegraphics[width=1.0\linewidth]{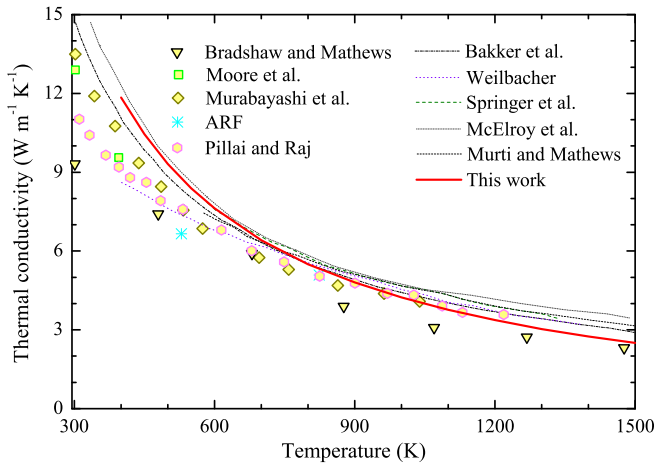}
\end{center}
\caption{(Color online) The thermal conductivity of ThO$_{2}$.
Experimental results from Springer \emph{et al.} \cite{Springer},
Weibacher \cite{Weilbacher}, McElroy \emph{et al.} \cite{McElroy},
ARF \cite{ARF}, Murabayashi \emph{et al.} \cite{Murabayashi}, Pillai
and Raj \cite{Pillai}, Bradshaw \cite{Bradshaw}, Moore \emph{et al.}
\cite{Moore}, Murti and Mathews \cite{Mufti}, and Bakker \emph{et
al.} \cite{Bakker} are displayed for comparison.} \label{5}
\end{figure}

\begin{figure}[ptb]
\begin{center}
\includegraphics[width=0.8\linewidth]{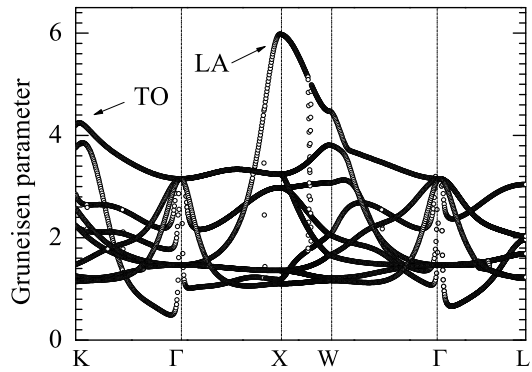}
\end{center}
\caption{(Color online) The mode Gr\"{u}neisen parameters along
high-symmetry directions in the reciprocal lattice space of
ThO$_{2}$.} \label{6}
\end{figure}

Based on Eq. (\ref{eq11}) and the obtained Debye temperature, we
calculate the thermal conductivity for ThO$_2$. Figure 5 shows our
thermal conductivity result, in comparison with the previous
experimental measurements in the temperature range from 300 to 1500
K. We can see that in the considered temperature range, our
calculated values are in agreement with the experimental results,
which proves the validity of our methods and model. Especially, in
the high temperature range from 600 to 1200 K, our results accords
very well with the experimental measurements by Murabayashi, {\it et
al.} \cite{Murabayashi} and by Pillai and Raj \cite{Pillai}. At the
low temperature range around the Debye temperature of 402 K, our
results are slightly different from some experimental results. This
difference comes from our presumption that the dominant mechanism
for phonon scattering is the Umklapp process. The accordance between
our calculations and corresponding experiments in the temperature
range from 600 to 1200 K proves that within this temperature area,
the contributions from other phonon-scattering mechanisms are so
small that can be neglected.

The mode Gr\"{u}neisen parameter describing the phonon frequency
shift with respect to the volume can be used to discuss the
anharmonic effects. By expanding or compressing the equilibrium
volume by 1\%, we calculate the mode Gr\"{u}neisen parameter
$\gamma_{j}(\mathbf{q})$ for all nine phonon branches according to
Eq. (\ref{eq5}). The corresponding results are shown in Fig. 6. It
can be seen that all the mode Gr\"{u}neisen parameter values are
positive, indicating that all phonon frequencies increase with
decreasing volume. Besides, the acoustic phonon mode Gr\"{u}neisen
parameters are relatively larger reflecting that changes in volume
have more influences on the collective vibration modes of ThO$_2$.
We can also see from Fig. 6 that the LA and TO phonon branches have
larger mode Gr\"{u}neisen parameters, indicating that the anharmonic
interactions between the LA and TO branches should be more intensive
with respect to the volume change.

\subsection{Defect formation in ThO$_2$}

\begin{table*}[ptb]
\caption{Formation energies of different defects at different charge
states in ThO$_{2}$. The defects include oxygen vacancy (V$_{\rm
O}$), interstitial oxygen ion (O$_{i}$), thorium vacancy (V$_{\rm
Th}$), interstitial thorium ion (Th$_{i}$), oxygen (FP$_{\rm O}$)
and thorium Frenkel-pairs (FP$_{\rm Th}$), and Schottky defect (S)
of a ThO$_2$ unit. The formation energies are in units of eV.}
\label{point}
\begin{ruledtabular}
\begin{tabular}{ccccccccccccc}
Defect&Charge on defect&Kr\"{o}ger-Vink notation&$E_{for}$\\
\hline
V$_{\rm O}$&0&V$_{\rm O}^{\rm X}$&7.415\\
V$_{\rm O}$&+1&V$_{\rm O}^{\bullet}$&3.922\\
V$_{\rm O}$&+2&V$_{\rm O}^{\bullet\bullet}$&1.338\\
O$_{i}$&0&O$_{i}^{\rm X}$&1.901\\
O$_{i}$&-1&O$_{i}^{'}$&4.026\\
O$_{i}$&-2&O$_{i}^{''}$&5.487\\
V$_{\rm Th}$&0&V$_{\rm Th}^{\rm X}$&18.349\\
V$_{\rm Th}$&-1&V$_{\rm Th}^{'}$&18.600\\
V$_{\rm Th}$&-2&V$_{\rm Th}^{''}$&18.314\\
V$_{\rm Th}$&-3&V$_{\rm Th}^{''''}$&18.327\\
V$_{\rm Th}$&-4&V$_{\rm Th}^{''''}$&18.462\\
Th$_{i}$&0&Th$_{i}^{\rm X}$&6.094\\
Th$_{i}$&+1&Th$_{i}^{\bullet}$&5.897\\
Th$_{i}$&+2&Th$_{i}^{\bullet\bullet}$&2.869\\
Th$_{i}$&+3&Th$_{i}^{\bullet\bullet\bullet}$&0.488\\
Th$_{i}$&+4&Th$_{i}^{\bullet\bullet\bullet\bullet}$&-1.741\\
FP$_{\rm O}$&0&V$_{\rm O}^{\rm X}$+O$_{i}^{\rm X}$&9.316\\
FP$_{\rm O}$&0&V$_{\rm O}^{\bullet}$+O$_{i}^{'}$&7.948\\
FP$_{\rm O}$&0&V$_{\rm O}^{\bullet\bullet}$+O$_{i}^{''}$&6.825\\
FP$_{\rm Th}$&0&V$_{\rm Th}^{\rm X}$+Th$_{i}^{\rm X}$&24.443\\
FP$_{\rm Th}$&0&V$_{\rm Th}^{'}$+Th$_{i}^{\bullet}$&24.497\\
FP$_{\rm Th}$&0&V$_{\rm Th}^{''}$+Th$_{i}^{\bullet\bullet}$&21.183\\
FP$_{\rm Th}$&0&V$_{\rm Th}^{'''}$+Th$_{i}^{\bullet\bullet\bullet}$&18.815\\
FP$_{\rm Th}$&0&V$_{\rm Th}^{''''}$+Th$_{i}^{\bullet\bullet\bullet\bullet}$&16.721\\
S&0&V$_{\rm Th}^{\rm X}$+2V$_{\rm O}^{\rm X}$&19.472\\
S&0&V$_{\rm Th}^{''}$+2V$_{\rm O}^{\bullet}$&12.651\\
S&0&V$_{\rm Th}^{''''}$+2V$_{\rm O}^{\bullet\bullet}$&8.231\\
\end{tabular}
\end{ruledtabular}
\end{table*}

In this subsection and the next, we will investigate the structural
stability of ThO$_2$ by systematically calculating the formation
energy of different kinds of defects and investigating the diffusion
behaviors of helium. A 2$\times$2$\times$2 supercell with 96 Th and
O atoms is employed in these two subsections to model defect
formation and helium diffusion in ThO$_2$. Different charge states
are considered for all the defects. To calculate the formation
energy, the positions of all ions are fully relaxed before we
calculate the electronic free energies of the system with different
defects.

\begin{figure*}[ptb]
\begin{center}
\includegraphics[width=0.8\linewidth]{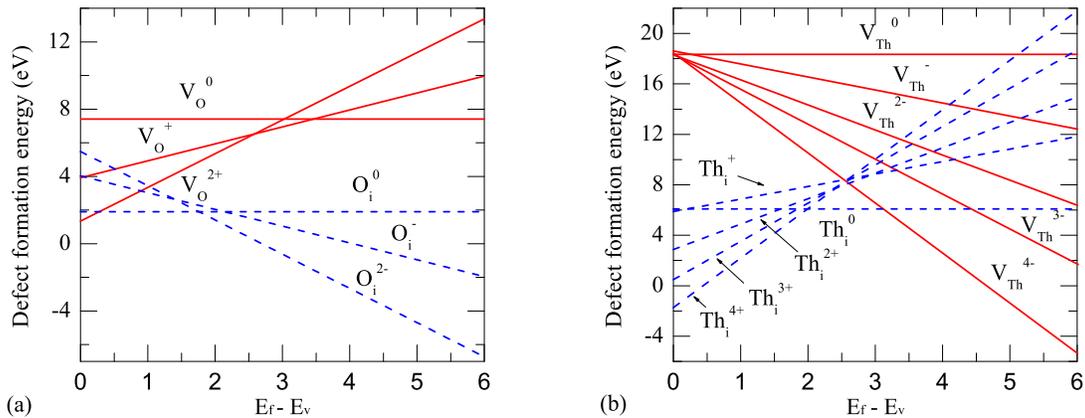}
\end{center}
\caption{(Color online) Formation energies of oxygen related
defects, i.e., oxygen vacancies or interstitial oxygen ions (a) and
thorium related defects, i.e., thorium vacancies or interstitial
thorium ions (b) as functions of the Fermi energy.} \label{7}
\end{figure*}

\begin{figure*}[ptb]
\begin{center}
\includegraphics[width=0.8\linewidth]{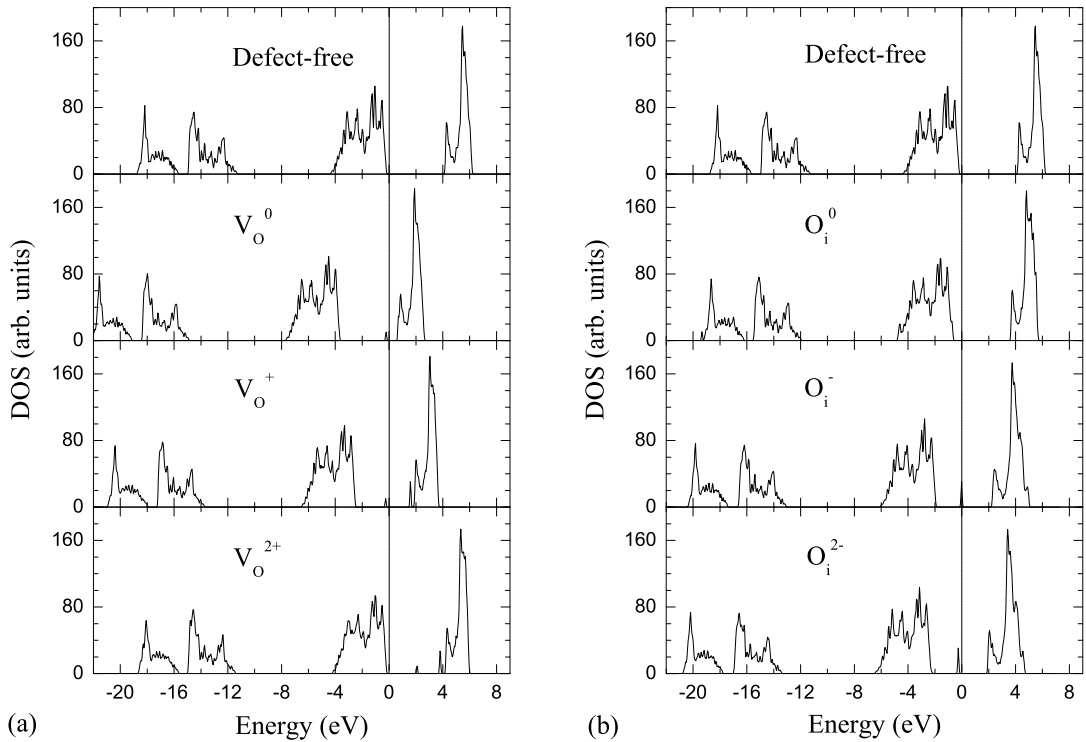}
\end{center}
\caption{(Color online) Electronic density of states (DOS) for
ThO$_2$ with oxygen vacancies (a), and with interstitial oxygen
defects (b) in different charge states. The Fermi energies are set
to zero.} \label{8}
\end{figure*}

\begin{figure*}[ptb]
\begin{center}
\includegraphics[width=0.8\linewidth]{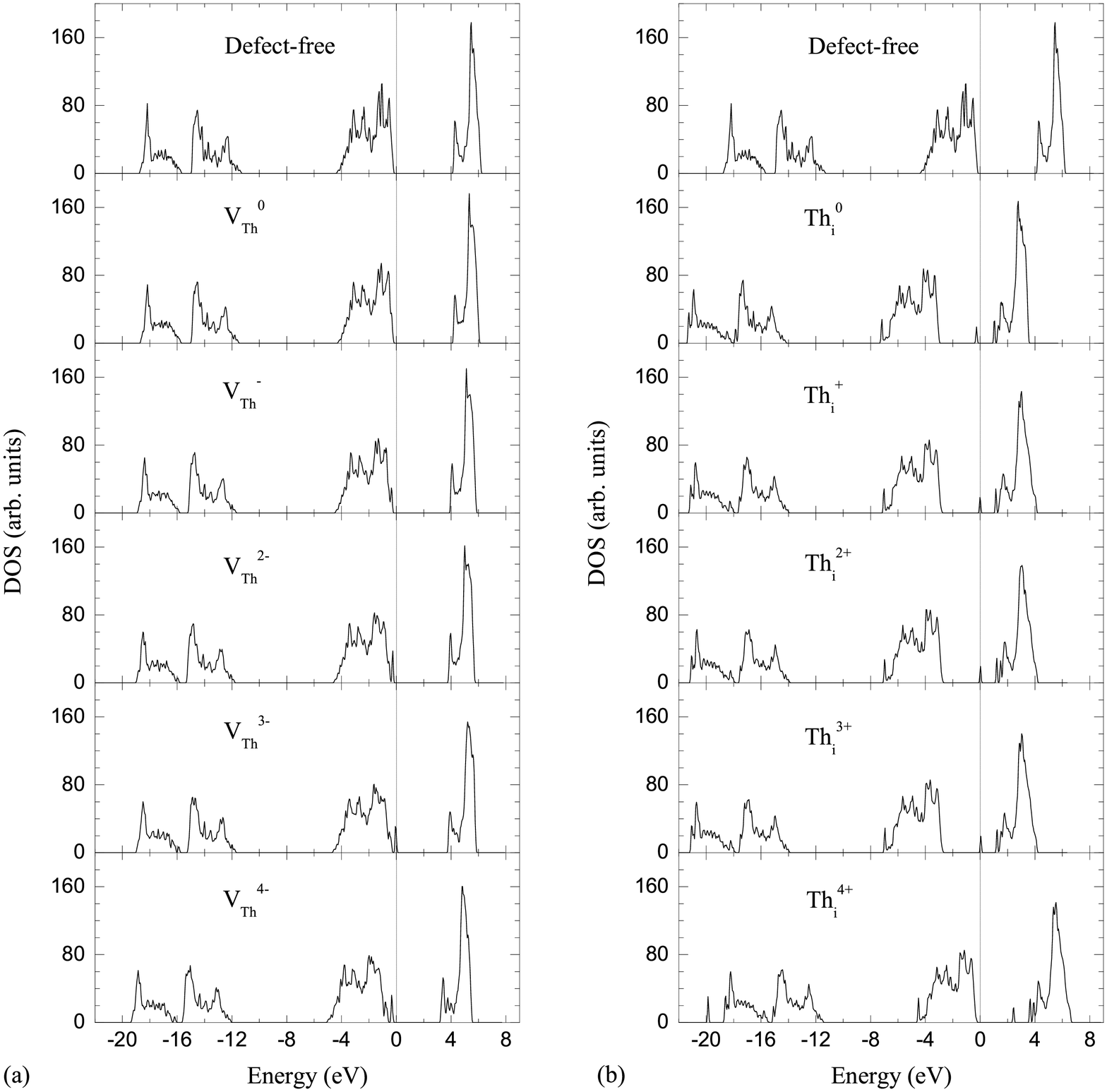}
\end{center}
\caption{(Color online) Electronic density of states (DOS) for
ThO$_2$ with thorium vacancies (a), and with interstitial thorium
defects (b) in different charge states. The Fermi energies are set
to zero.} \label{9}
\end{figure*}

The formation energy for oxygen (thorium) vacancies, interstitial
oxygen (thorium) ions, oxygen (thorium) Frenkel pairs, and the
Schottky defect of a ThO$_2$ unit in the 2$\times$2$\times$2
supercell are obtained and listed in Table \ref{point}. One can see
that the charged Th$_i^{4+}$ add-in defect is the only one having a
negative formation energy. It means that once Th$_i^{4+}$ ions are
available, interstitial Th$_i^{4+}$ defect can form in ThO$_2$ with
an energy release of 1.74 eV. Comparatively for oxygen defects, the
O$^{2-}$ vacancy is the most possible one because of its smallest
formation energy among oxygen defects. These two results reflects
the ionic character of the Th-O bonds that each thorium atom almost
loses 4 electrons to two oxygen atoms in ThO$_2$. From Table
\ref{point}, we can also see that all thorium vacancies and thorium
Frenkel-pairs have huge formation energies and thus can hardly form
in intrinsic ThO$_2$. Comparatively, the formation of oxygen
Frenkel-pair, containing an O$^{2+}$ vacancy and a interstitial
O$^{2-}$ ion has a relatively smaller formation energy.

Depending on the environment's influence on the Fermi energy of the
ThO$_2$ material, the formation possibility of charged defects can
change due to the changes on hardness of electrons transfer. Using
Eqs. (7)-(9) under different Fermi energies, we can calculate the
formation energy as a function of the Fermi energy for different
kinds of defects. The corresponding results for oxygen (thorium)
vacancies and interstitial oxygen (thorium) ions in different charge
states are respectively shown in Figs. 7(a) and 7(b). The
experimentally measured band-gap of 6.0 eV, instead of the 4.1 eV
value obtained by first-principles calculations is chosen as the
reference band-gap in discussions on defect formation.

As can be seen from Fig. 7(a), when comparing the energies
associated with forming V$_{\mathrm{O}}^{0}$, V$_{\mathrm{O}}^{+}$,
and V$_{\mathrm{O}}^{2+}$ oxygen vacancies, a transition can be
observed with the Fermi level increasing from the valence band to
the conduction band. The +2 charged oxygen vacancy is favored near
the valence band, indicating that oxygen vacancies have a tendency
to donate electrons or behave as a $n$-type defect. When the Fermi
level increases to around 2.6 eV, the V$_{\mathrm{O}}^{+}$ becomes
energetically favorable. With further increasing the Fermi level,
the neutral charge state is most probable for oxygen vacancies, and
the tendency of oxygen vacancy to donate electrons diminishes. We
can also see from Fig. 7(a) that the interstitial O$_{i}^{2-}$
defect can become very possible to form when the Fermi energy is
shifted to be near the conduction band of ThO$_2$.

From Fig. 7(b) we see that when the Fermi energy is near the valence
band, all charged states of thorium vacancy are hard to form because
of the huge formation energies, and the interstitial Th$_{i}^{4+}$
ions can easily form in ThO$_2$ with a negative formation energy.
With upshifting the Fermi energy, the interstitial thorium defect in
neutral state and the V$_{\rm Th}^{4-}$ vacancy can both become the
most possible thorium kinds of defects. When considering only the
vacancy defects, V$_{\mathrm{O}}^{2+}$ is the most stable defect
near the valence band, while near the conduction band,
V$_{\mathrm{Th}}^{4-}$ is the most favorable one. Comparatively for
interstitial states, the Th$_{i}^{4+}$ and O$_{i}^{2-}$ defects are
the most probable ones when the Fermi energy is near the valence and
conduction band respectively.

In addition, the electronic density of states (DOS) for both
defect-free and defective ThO$_{2}$ are calculated and shown in
Figs. 8 and 9, to further analyze the influences of the considered
defects on the electronic structures of ThO$_2$. As clearly shown in
both Figs. 8 and 9, the introduction of vacancies or interstitial
ions do not change the DOS distribution of ThO$_2$ very much. The
biggest character in both Figs. 8 and 9 is that a new defect energy
level emerges in the band gap for defective ThO$_2$. For the
electronic structure of ThO$_2$ with oxygen vacancies, we see from
Fig. 8(a) that the Fermi energy shifts from the valence band maximum
(VBM) in defect-free ThO$_2$, to above the defect energy level for
the V$_{\rm O}^{0}$ defect, and from above the defect energy level
for the V$_{\rm O}^{+}$ defect back to the VBM for the V$_{\rm
O}^{2+}$ defect. For the system with interstitial oxygen ions, we
can see from Fig. 8(b) that the Fermi energy shifts respectively
from the VBM for the O$_{i}^{0}$ defect to the defect energy level
for the O$_{i}^{-}$ defect, and from the defect energy level for the
O$_{i}^{-}$ defect to above the defect energy level for the
O$_{i}^{2-}$ defect. Similarly for the ThO$_2$ supercell with
interstitial thorium ions, the Fermi energy shifts from the VBM to
the defect energy level for the Th$_{i}^{+}$, Th$_{i}^{2+}$, and
Th$_{i}^{3+}$ defects, and then shifts to above the defect energy
level for the Th$_{i}^{4+}$ defect, as shown in Fig. 9(b). From Fig.
9(a), one can see that the defect energy levels for thorium
vacancies in ThO$_2$ are very close to the VBM.

\subsection{Diffusion of helium in ThO$_2$}

\begin{figure*}[ptb]
\begin{center}
\includegraphics[width=1.0\linewidth]{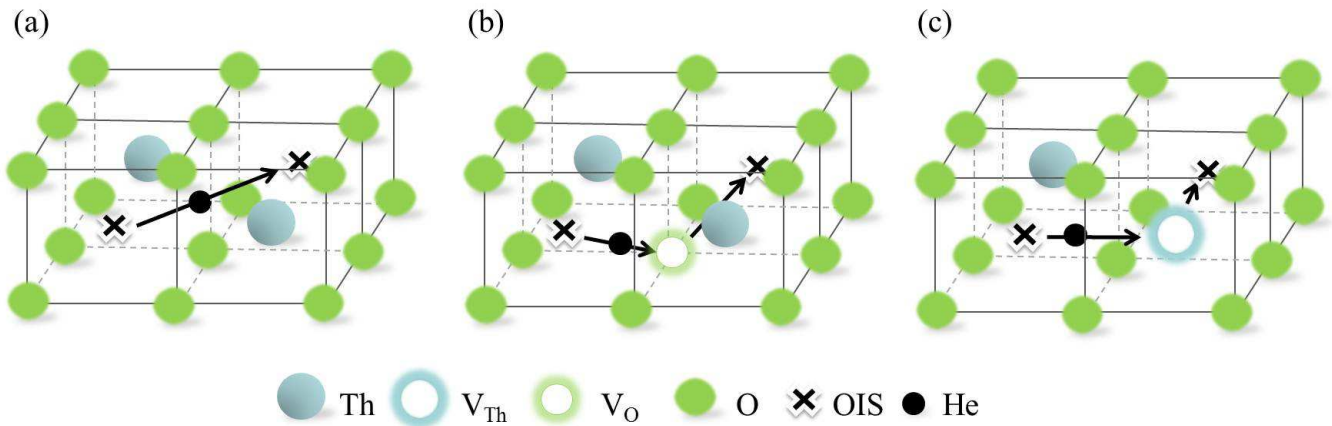}
\end{center}
\caption{(Color online) Diffusion pathways for a helium atom from
one octahedral interstitial site to another octahedral interstitial
site directly in intrinsic ThO$_2$ (a), through an oxygen vacancy
(b) and through a thorium vacancy in defective ThO$_2$ (c).}
\label{10}
\end{figure*}

\begin{figure*}[ptb]
\begin{center}
\includegraphics[width=1.0\linewidth]{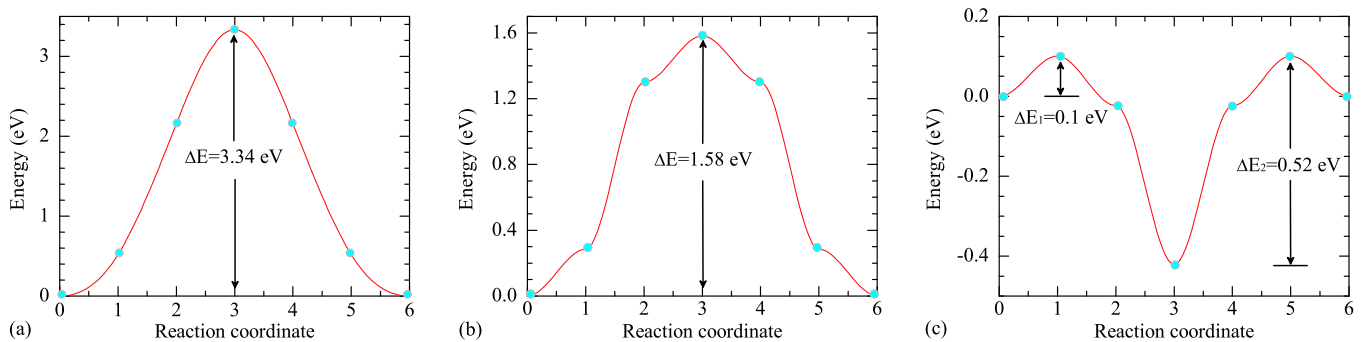}
\end{center}
\caption{(Color online) The energy profiles for a helium atom to
diffuse from one octahedral interstitial site to another octahedral
interstitial site directly in intrinsic ThO$_2$ (a), through an
oxygen vacancy (b) and through a thorium vacancy in defective
ThO$_2$ (c).} \label{11}
\end{figure*}

In order to determine the structure influence of helium impurity on
ThO$_{2}$, we systematically calculate the incorporation energy and
diffusion energy barriers for helium in both the defect-free and
defective ThO$_2$. The incorporation energy is defined to be the
energy required to incorporate one helium atom at a pre-existing
vacancy or at an interstitial site. In this way, $E_{inc}$ can be
expressed as follows
\begin{equation}
E_{inc}=E_{tot}-E_{\rm ThO_2}-E_{\rm He},
\end{equation}
where $E_{tot}$ is the energy of the ThO$_{2}$ supercell with an
incorporated helium, $E_{\rm ThO_2}$ is the energy of the ThO$_2$
supercell with or without defects, and $E_{\rm He}$ the energy of an
isolated helium atom. Three different kinds of incorporations are
systematically considered here, i.e., a helium atom at the oxygen
(O) and thorium vacancy (Th) sites in defective ThO$_2$, and a
helium atom at the octahedral interstitial site (OIS). The
calculated incorporation energies are 2.48, 0.21, and 0.77 eV
respectively. We can see that the most possible incorporation site
for helium is the thorium vacancy site in defective ThO$_2$. In
comparison, the oxygen vacancy site is much less possible for helium
incorporation.

The structural influence of helium on ThO$_2$ is determined not only
by the incorporation energy, but also by the minimum energy path for
helium diffusion. The climbing image nudged elastic band (CINEB)
method \cite{Henkelman} is then employed to find the minimum-energy
diffusion pathways. Based on the obtained incorporation results for
helium in ThO$_2$, we here investigate the diffusions of helium from
OIS to OIS in both defect-free and defective ThO$_2$. Figures
10(a)-(c) depict our considered diffusion ways. And the energy
profiles along these paths are shown in Figs. 11(a)-(c)
respectively. We can see from Fig. 11(a) that in intrinsic ThO$_2$,
the incorporated helium atom needs to overcome a 3.34 eV energy
barrier in order to diffuse from one OIS to another. This huge
diffusion energy barrier means that at normal temperatures, helium
diffusion in intrinsic ThO$_2$ is almost impossible. In the
corresponding saddle point along this diffusion path, the helium
atom is at the middle of the two OISs, and the nearest oxygen atoms
along the [001] direction are repelled from each other by about 0.80
\AA.

From the calculated energy profile shown in Fig. 11(b), we see that
the energy barrier for an incorporated helium atom to diffuse from
one OIS to another by passing an oxygen vacancy is lowered down to
be 1.58 eV. This energy barrier is however still too large for the
diffusion to happen at room temperatures. The only possible way for
helium to diffuse in ThO$_2$ is found to be passing through a
thorium vacancy, as depicted in Fig. 10(c). We can see from Fig.
11(c) that the helium atom only needs to overcome a 0.10 eV energy
barrier to diffuse from an OIS to the thorium vacancy, and a 0.52 eV
energy barrier to diffuse from the thorium vacancy to another OIS.
The small energy barriers for helium to diffuse in thorium
vacancy-included ThO$_2$ indicate that thorium vacancies might lead
to helium aggregation causing failure of the ThO$_2$ material.
Fortunately from our above defect formation studies, the formation
of thorium vacancies is almost forbidden when the Fermi energy is
not upshifted, because of their too large formation energy in
ThO$_2$. Therefore, to keep ThO$_2$ away from structural damages
from helium incorporation, any kinds of factors possibly leading to
upshifts of the Fermi energy should be avoided.

\section{Summary}

In summary, we have performed a systematic first-principles study to
investigate the thermodynamic properties and structural stabilities
of ThO$_{2}$. Based on the calculated phonon dispersion curves for
ThO$_2$, we systematically analyze its thermodynamic properties and
obtain the values of its thermal expansion coefficient, bulk
modulus, and heat capacities at different temperatures, which are in
good agreement with corresponding experimental measurements. The
agreement between our calculations and experiments also proves the
validity of our methods and model, and the effectiveness of the
quasi-harmonic approximation. According to the Umklapp interaction
mechanism between different phonon branches, we systematically
obtain the mode Gr\"{u}neisen parameters, and further calculate the
thermal conductivities of ThO$_2$. Within the temperature range from
Debye temperature to about 1500 K, our calculated thermal
conductivity accords very well with experimental results.

In addition to studying the thermodynamic properties of ThO$_2$, we
also investigate its structural stability by calculating the
formation energy of different defects, and the diffusion behaviors
of helium, during which different charge states of the defects are
considered. The formation energy results indicate that without any
shifts of the Fermi energy, the interstitial Th$^{4+}$ defect is
very probable to appear in ThO$_2$ with an energy release of 1.74
eV. With changing the Fermi energy to different values, the
formation possibilities of different defects varies. For helium
incorporation, it is found that the helium atom tends to occupy a
thorium vacancy in defective ThO$_2$ or occupy the octahedral
interstitial site in intrinsic ThO$_2$. It is further revealed that
incorporated helium atoms can only diffuse freely in the
thorium-vacancy contained ThO$_2$, with small energy barriers of
0.10 and 0.52 eV. Our studies point out that to avoid helium damage,
the electronic Fermi energy of ThO$_2$ should not be upshifted
because it can makes the formation of thorium vacancies less
possible.

\section{\label{sec:level1}ACKNOWLEDGMENTS}
This work was supported by NSFC under Grant No. 51071032, and by
Foundations for Development of Science and Technology of China
Academy of Engineering Physics under Grants No. 2011A0301016 and No.
2011B0301060.


\begin{thebibliography}{00}

\bibitem{Belle69}
J. Belle, J. Nucl. Mater. {\bf 30}, 3 (1969).

\bibitem{Killeen80}
J. C. Killeen, J. Nucl. Mater. {\bf 88}, 185 (1980).

\bibitem{Kudin02}
K. N. Kudin, G. E. Scuseria, and R. L. Martin, Phys. Rev. Lett. {\bf
89}, 266402 (2002).

\bibitem{Dorado09}
B. Dorado, B. Amadon, M. Freyss, and M. Bertolus, Phys. Rev. B {\bf
79}, 235125 (2009).

\bibitem{Dorado10}
B. Dorado, G. Jomard, M. Freyss, and M. Bertolus, Phys. Rev. B {\bf
82}, 035114 (2010).

\bibitem{Petit10}
L. Petit, A. Svane, Z. Szotek, W. M. Temmerman, and G. M. Stocks,
Phys. Rev. B {\bf 81}, 045108 (2010).

\bibitem{Herring}
J. S. Herring, P. E. MacDonald, K. D. Weaver, C. Kullberg, Nucl.
Eng. Des. {\bf 203}, 65-85 (2001).

\bibitem{Tommasi}
J. Tommasi, A. Puill, and Y.K. Lee, "Reactors with
Th/Pu Fuels," Proc. Workshop Advanced Reactors with Innovative
Fuels, Villigen, Switzerland, October 21-23 (1998).

\bibitem{Maitura}
R. Maitura, Nucl. Rep. 46-53 (2005).

\bibitem{Chang}
H. Chang, Y. Yang, X. Jing, and Y. Xu, "Thorium-Based Fuel
Cycles in the Modular High Temperature Reactor," Tsinghua Sci. Tech.
{\bf 11}, 6 (2006).

\bibitem{Carrera}
A. Carrera, J. Lacouture, C. Campo, and G. Paredes, "Feasibility
Study of Boiling Water Reactor Core Based on Thorium-Uranium Fuel
Concept," Energy Conversion Management, {\bf 49}, 1 (2007).

\bibitem{Juhn}
P. E. Juhn, "Thorium Fuel Cycle Options for Advanced Reactors:
Overview of IAEA Activities," Proc. Workshop Advanced Reactors with
Innovative Fuels, Villigen, Switzerland, October 21-23 (1998).

\bibitem{Momin}
A. C. Momin and Karkhanvala, High Temp. Sci. {\bf 10}, 45 (1978).

\bibitem{Springer}
J. R. Springer, E. A. Eldridge, M. U. Goodyear, T. R. Wright and J.
F. Lagedrast Report No. BMI-X-10210 (1968).

\bibitem{Weilbacher}
J. C. Weilbacher, High Temp.-High Press. {\bf 4}, 431 (1972).

\bibitem{McElroy}
D. L. McElroy, J. P. Moore, P. H. Spindler, Oak Ridge National
Laboratory Report ORNL-4429, p. 121 (1968).

\bibitem{ARF}
Armour Research Foundation ARF-Project No. 6-025, Final Report,
(1957).

\bibitem{Murabayashi}
M. Murabayashi, J. Nucl. Sci. Technol. {\bf 7}, 559 (1970).

\bibitem{Pillai}
C. G. S. Pillai, P. Raj, J. Nucl. Mater. {\bf 277}, 116-119 (2000)

\bibitem{Bradshaw}
W. G. Bradshaw, C. O. Mathews, Report LMSD-2466, (1958).

\bibitem{Moore}
J. P. Moore, R. S. Graves, T. G. Kollie, D. L. McEIroy, Oak Ridge
National Laboratory Report ORNL-4121, (1967).

\bibitem{Mufti}
P. Srirama Mufti, C. K. Mathews, J. Phys. D {\bf 24}, 2202 (1991).

\bibitem{Bakker}
K. Bakker, E. H. P. Corfunke, R. J. M. Konings and R. P. C. Scharm,
J. Nucl. Mater. {\bf 250}, 1-12 (1997).

\bibitem{Chadwick}
D. Chadwick and J. Graham, Nat. Phys. Sci. {\bf 237}, 127 (1972).

\bibitem{Allen}
G. C. Allen and P. M. Tucker, J. Chem. Soc. Dalton p. 470 (1973).

\bibitem{Veal}
B. W. Veal and D. J. Lam, Phys. Rev. B {\bf 10}, 12 (1974).

\bibitem{Jayaraman}
A. Jayaraman, G. A. Kourouklis, L. G. Van Uitert, Pramana {\bf 30},
225 (1988).

\bibitem{Dancausse}
J. P. Dancausse, E. Gering, S. Heathman, U. Benedict, High Pressure
Res. {\bf 2}, 381 (1990).

\bibitem{Idiri}
M. Idiri, T. Le Bihan, S. Heathman, J. Rebizant, Phys. Rev. B {\bf
70}, 014113 (2004).

\bibitem{Kim}
Y. -E. Kim, J. -W. Park and J. Cleveland, "Thermophysical properties database of materials for light water reactors and heavy water
reactors", Vienna, (2006).

\bibitem{Sevik}
C. Sevik and T. Cagin, Phys. Rev. B {\bf 80}, 014108 (2009).

\bibitem{Kanchana}
V. Kanchana, G. Vaitheeswaran, A. Svane, A. Delin, J. Phys.:
Condens. Matter {\bf 18}, 9615 (2006).

\bibitem{Shein}
I. R. Shein, K. I. Shein, A. L. Ivanovskii, J. Nucl. Mater. {\bf
361}, 69 (2007).

\bibitem{Boettger}
J. C. Boettger, Int. J. Quantum. Chem. {\bf 109}, 3564 (2009).

\bibitem{Wang}
B. T. Wang, H. L. Shi , W. D. Li , P. Zhang, J. Nucl. Mater. {\bf
399}, 181-188 (2010).

\bibitem{Boudjemline}
A. Boudjemline, L. Louail, Mazharul M. Islam, B. Diawara, Comp.
Mater. Sci. {\bf 50}, 2280-2286 (2011).

\bibitem{G.Kresse1}
G. Kresse, J. Furthm\"{u}ller, computer code VASP, Vienna, (2005).

\bibitem{G.Kresse2}
G. Kresse, J. Furthm\"{u}ller, Phys. Rev. B {\bf 54}, 11169 (1996).

\bibitem {PAW}
P. E. Bl\"{o}chl, Phys. Rev. B {\bf 50}, 17953 (1994).

\bibitem{PBE}
J. P. Perdew, K. Burke, M. Ernzerhof, Phys. Rev. Lett. {\bf 77},
3865 (1996).

\bibitem{Monkhorst}
H. J. Monkhorst, J. D. Pack, Phys. Rev. B {\bf 13}, 5188 (1976).

\bibitem{Baldereschi}
A. Baldereschi, S. Baroni, R. Resta, Phys. Rev. Lett. {\bf
61}, 734 (1988).

\bibitem{Peressi}
M. Peressi, N. Binggeli, A. Baldereschi, J. Phys. D {\bf 31}, 1273 (1998).

\bibitem{Brich}
F. Brich, Phys. Rev. {\bf 71}, 809 (1947).

\bibitem{Olsen}
J. S. Olsen, L. Gerward, V. Kanchana, G. Vaitheeswaran, J. Alloys
Compd. {\bf 381}, 37 (2004).

\bibitem{Lu}
Y. Lu, D. F. Li, R. W. Li, H. L. Shi, P. Zhang, J. Nucl. Mater. {\bf
408}, 136-141 (2011).

\bibitem{Zhang10}
P. Zhang, B. T. Wang, and X. G. Zhao, Phys. Rev. B {\bf 82}, 144110
(2010).

\bibitem{Macedo}
P. M. Macedo, W. Capps, J. B. Watchman, J. Am. Ceram. Soc. {\bf 47},
651 (1964).

\bibitem{Touloukian}
Y. S. Touloukian, R. K. Kirby, R. E. Taylor, T. Y. R. Lee, Thermal
Expansion. Nonmetallic Solids (IFI/Plenum, New York, 1970).

\bibitem{Slack}
G. A. Slack, J. Phys. Chem. Solids {\bf 34}, 321 (1973).

\bibitem{Slack1979}
G. A. Slack, Solid State Phys. {\bf 34}, 1 (1979).

\bibitem{Anderson}
O. L. Anderson, J. Phys. Chem. Solids {\bf 12}, 41 (1959).

\bibitem{Blanco}
A. A. Blanco, E. Francisco, and V. Luana, Comput. Phys. Commun. {\bf
158}, 57 (2004).

\bibitem{Francisco}
E. Francisco, J. M. Recio, M. A. Blanco, A. Mart\'{\i}n Pend\'{a}s,
J. Phys. Chem. {\bf 102}, 1595 (1998).

\bibitem{Francisco2}
E. Francisco, M. A. Blanco, and G. Sanjurjo, Phys. Rev. B {\bf 63},
094107 (2001).

\bibitem{Henkelman}
G. Henkelman, B. P. Uberuaga, and H. Jonsson, J. Chem. Phys. {\bf
113}, 9901 (2000).

\end{thebibliography}
\end{document}